\documentclass[11pt]{article}  
\usepackage{menuproc}
\usepackage{cite}
\usepackage{epsfig}
\usepackage{amsmath,amssymb}
\begin{document}
%
%
%
\titlematter{Baryon Spectroscopy at Beijing Electron Positron Collider}%
{B.S.Zou, X.B.Ji and H.X.Yang,  representing BES Collaboration}%
{Institute of High Energy Physics, CAS,\\
     P.O.Box 918 (4), Beijing 100039, P.R.China\\
 }
{BES Collaboration has collected 58 million $J/\psi$ events at the 
Beijing Electron-Positron Collider (BEPC). $J/\psi$ decays provide
an excellent place for studying excited nucleons and hyperons -- $N^*$,
$\Lambda^*$, $\Sigma^*$ and $\Xi^*$ resonances. 
Physics motivation, data status, partial wave analyses and
future prospects are presented for the baryon resonance program at BES.}
%

\section{Physics Motivation}

Baryons are the basic building blocks of our world. If we cut any piece
of object smaller and smaller, we will finally reach the nucleons, 
{\sl i.e.}, the lightest baryons, and we cannot cut them smaller any
further. So without mention any theory, we know that the study of baryon
structure is at the forefront of exploring microscopic structure of
matter.  From theoretical point of view,  since baryons represent the
simplest system in which the three colors of QCD neutralize into
colorless objects and the essential non-Abelian character of QCD is
manifest, understanding the baryon structure is absolutely necessary
before we claim that we really understand QCD.

Spectroscopy has long proved to be a powerful tool for exploring internal
structures and basic interactions of microscopic world. Ninety years ago
detailed studies of atomic spectroscopy resulted in the great
discovery of Niels Bohr's atomic quantum theory\cite{Bohr}. Forty to sixty
years later, still detailed studies of nuclear spectroscopy resulted in
Nobel Prize winning discoveries of nuclear shell model\cite{SM} and  
collective motion model\cite{CM} by Aage Bohr {\sl et al}.
Comparing with the atomic and nuclear spectroscopy at those times,
our present baryon spectroscopy is still in  its infancy\cite{PDG}.
Many fundamental issues in baryon spectroscopy are still not well
understood\cite{Capstick1}. The possibility of new, as yet unappreciated,
symmetries could be addressed with accumulation of more data. The new 
symmetries may not have obvious relation with QCD, just like nuclear
shell model and collective motion model.

\begin{figure}[htbp]
\vspace{-0.8cm}
\hspace{1.5cm}\includegraphics[width=14cm,height=6cm]{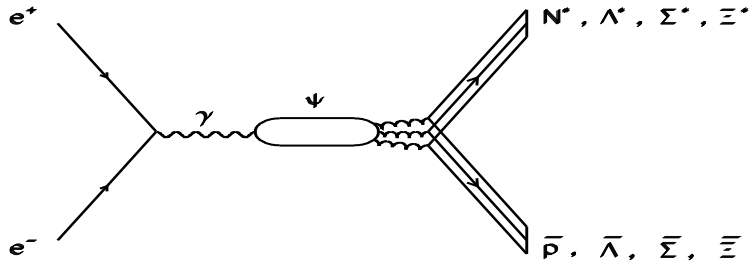}
\vspace{-1.1cm}
\caption{$\bar pN^*$, $\bar\Lambda\Lambda^*$,
$\bar\Sigma\Sigma^*$ and $\bar\Xi\Xi^*$ production
from $e^+e^-$ collision through $\Psi$ meson.}
\label{fig:1}
\end{figure}

Joining the new effort on studying the excited nucleons, $N^*$ baryons,   
at new facilities such as CEBAF at JLAB, ELSA at Bonn, GRAAL at Grenoble
and SPRING8 at JASRI, we also started a baryon resonance program at    
BES\cite{Zou1}, at Beijing Electron-Positron Collider (BEPC).
The $J/\psi$ and $\psi'$ experiments at BES provide an
excellent place for studying excited nucleons and hyperons -- $N^*$,
$\Lambda^*$, $\Sigma^*$ and $\Xi^*$ resonances\cite{Zou2}.
The corresponding Feynman graph for the production of these excited
nucleons and hyperons is shown in Fig.~\ref{fig:1} where $\Psi$
represents either $J/\psi$ or $\psi'$.

Comparing with other facilities, our baryon program has advantages in at 
least three obvious aspects:

(1) We have pure isospin 1/2 $\pi N$ and $\pi\pi N$ systems from
$J/\psi\to\bar NN\pi$ and $\bar NN\pi\pi$ processes due to isospin
conservation, while $\pi N$ and $\pi\pi N$ systems from $\pi N$ and
$\gamma N$ experiments are mixture of isospin 1/2 and 3/2, and suffer
difficulty on the isospin decomposition;

(2) $\psi$ mesons decay to baryon-antibaryon pairs through three or
more gluons. It is a favorable place for producing hybrid (qqqg) baryons,
and for looking for some``missing" $N^*$ resonances which have weak
coupling to both $\pi N$ and $\gamma N$, but stronger coupling to $g^3N$;

(3) Not only $N^*$, $\Lambda^*$, $\Sigma^*$ baryons,
but also $\Xi^*$ baryons with two strange quarks can be studied.   
Many QCD-inspired models\cite{Isgur,Glozman} are expected to be more 
reliable for baryons with two strange quarks due to their heavier quark
mass. More than thirty $\Xi^*$ resonances are predicted where only
two such states are well established by experiments. The theory is totally
not challenged due to lack of data.

\section{Data Status}\label{sec:form}

The BEijing Spectrometer (BES) is a conventional solenoidal magnet
detector that is described in detail in Ref.~\cite{bes}. A four-layer
central drift chamber(CDC) surrounding the beampipe provides trigger
information. A forty-layer cylindrical main drift chamber(MDC), located
radially outside the CDC, provides trajectory and energy loss ($dE/dX$)
information for charged tracks over $85\%$ of the total solid angle. 
An array of 48 scintillation counters surrounding the MDC measures the
time-of-flight(TOF) of charged tracks.
Radially outside of TOF system is a 12 radiation length thick,
lead-gas barrel shower counter(BSC) operating in the limited streamer
mode. This device covers $\sim 80\%$ of the total solid angle and measures
the energies of electrons and photons.

BES started data-taking in 1989 and was
upgraded in 1998. The upgraded BES is named BESII while the previous one
is called BESI. BESI collected 7.8 million $J/\psi$ events and 3.7 million
$\psi'$ events. BESII has collected 50 million $J/\psi$ events.

\begin{figure}[htbp]
\centerline{ \psfig{file=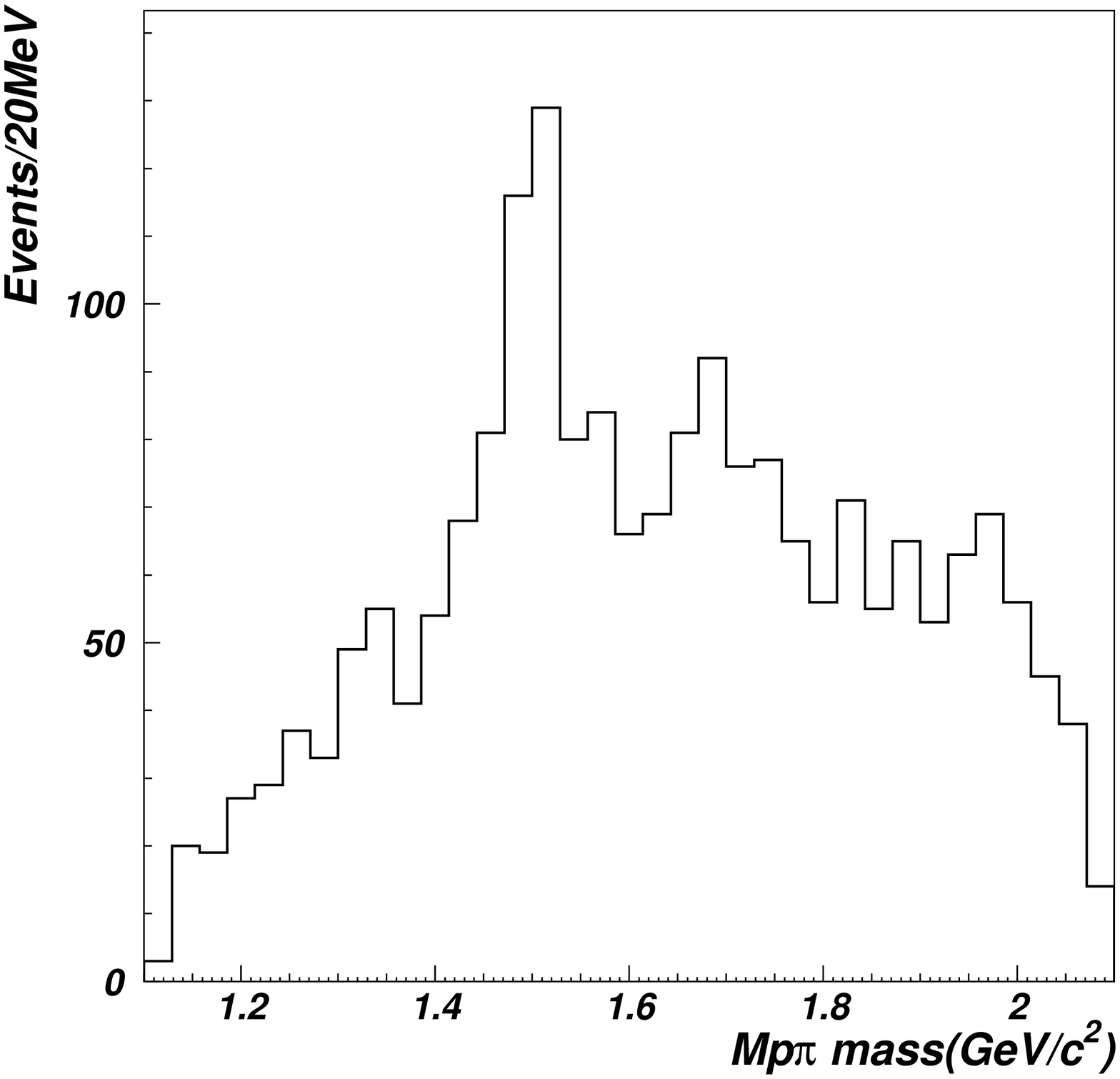,height=7.cm,angle=0,silent=}
\hspace{1.0cm} \psfig{file=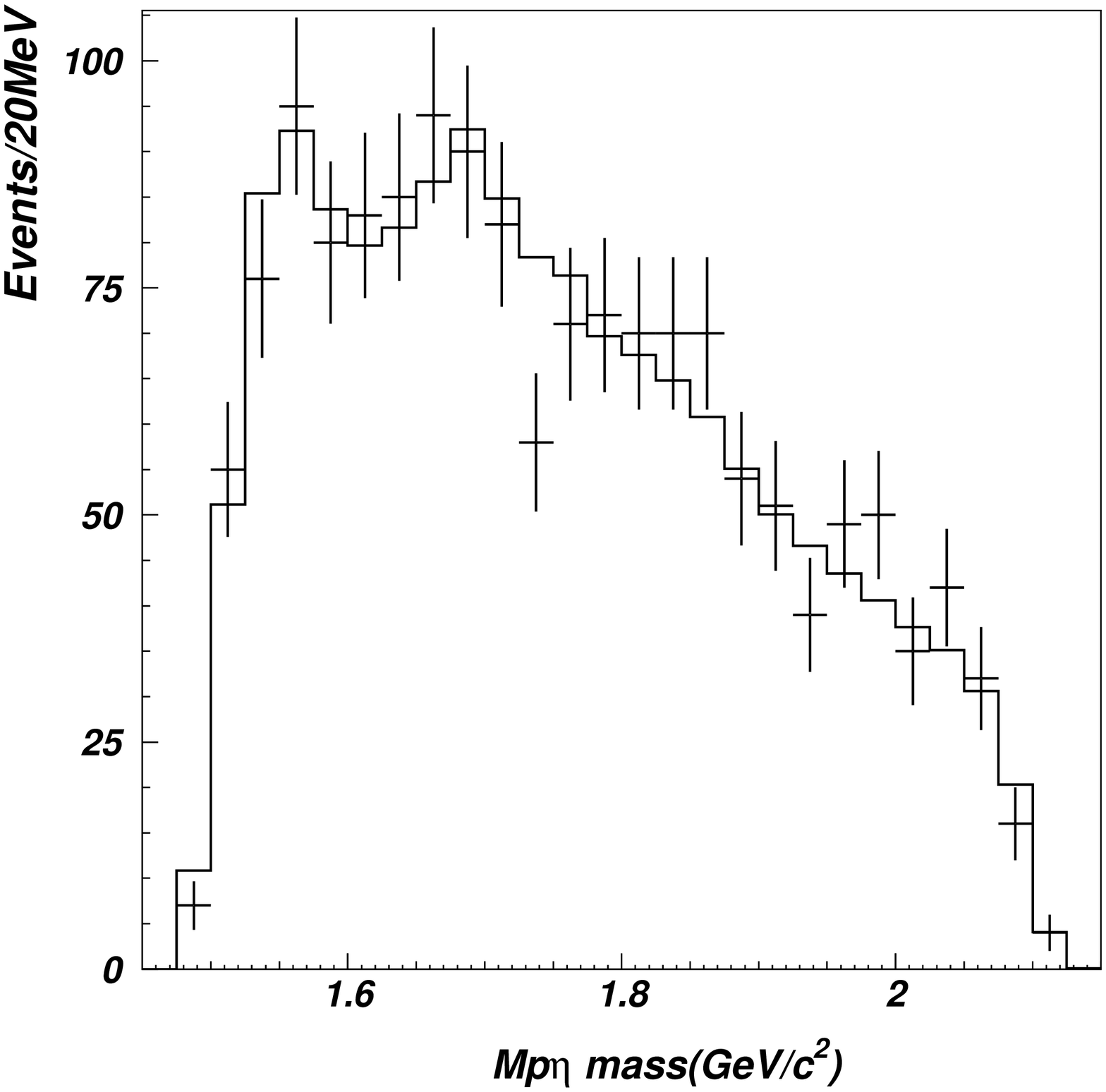,height=7.cm,angle=0,silent=}
}
\caption{\label{fig2} left: $p\pi^0$ invariant mass spectrum for   
$J/\psi\to\bar pp\pi^0$; right: $p\eta$ invariant mass spectrum for
$J/\psi\to\bar pp\eta$. BESI data}
\end{figure}

Based on 7.8 million $J/\psi$ events collected at BESI before 1996,
the events for $J/\Psi\to\bar pp\pi^0$ and $\bar pp\eta$ have been 
selected and reconstructed with $\pi^0$ and $\eta$ detected in their
$\gamma\gamma$ decay mode\cite{Zou1}.
The corresponding $p\pi^0$ and $p\eta$ invariant mass spectra are
shown in Fig.~\ref{fig2} with clear peaks around 1500 and
1670 MeV for $p\pi^0$ and clear enhancement around the $p\eta$ threshold,
peaks at 1540 and 1650 MeV for $p\eta$.

\begin{figure}[htbp]
\centerline{ \psfig{file=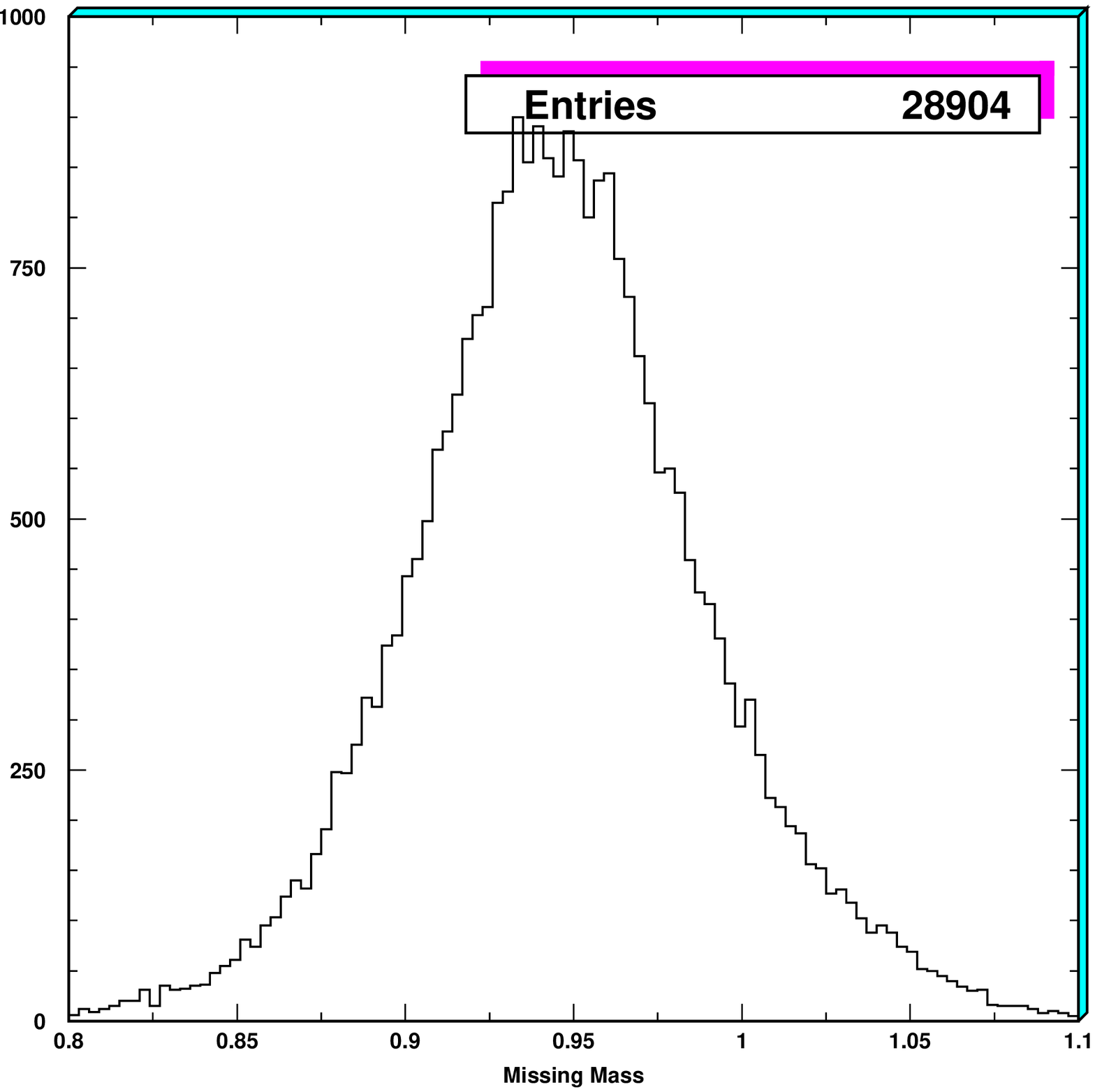,height=7.cm,angle=0,silent=}
\hspace{1.0cm} \psfig{file=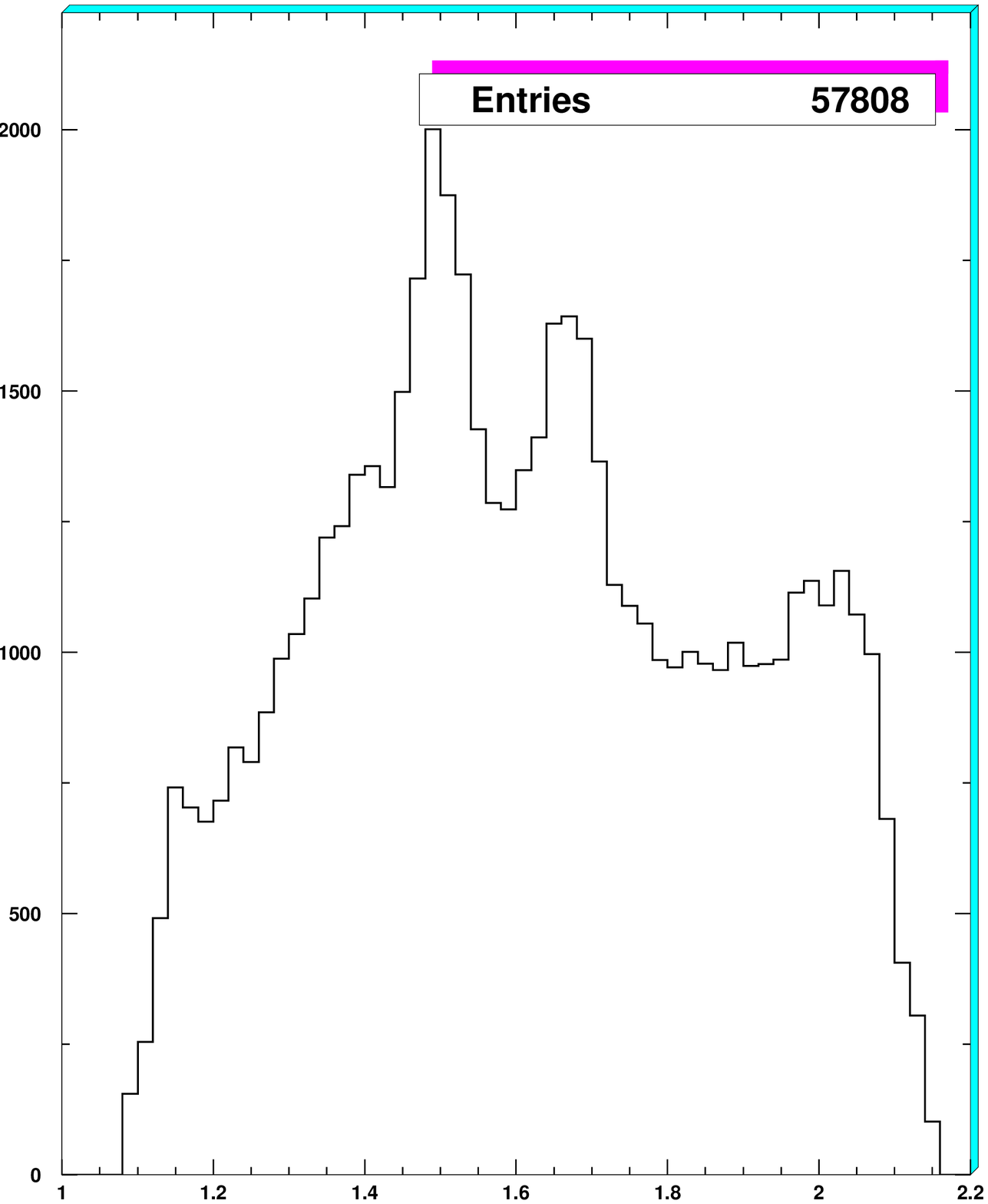,height=7.cm,
width=7.cm,angle=0,silent=}
}
\caption{\label{fig3} left: missing mass spectrum against $p\pi^-$ for   
$J/\psi\to\bar np\pi^-$; right: $p\pi^-\&\bar n\pi^-$ invariant mass
spectrum for
$J/\psi\to\bar np\pi^-$. Preliminary BESII data}
\end{figure}

\begin{figure}[htbp]
\vspace{1.2cm}
\centerline{ \psfig{file=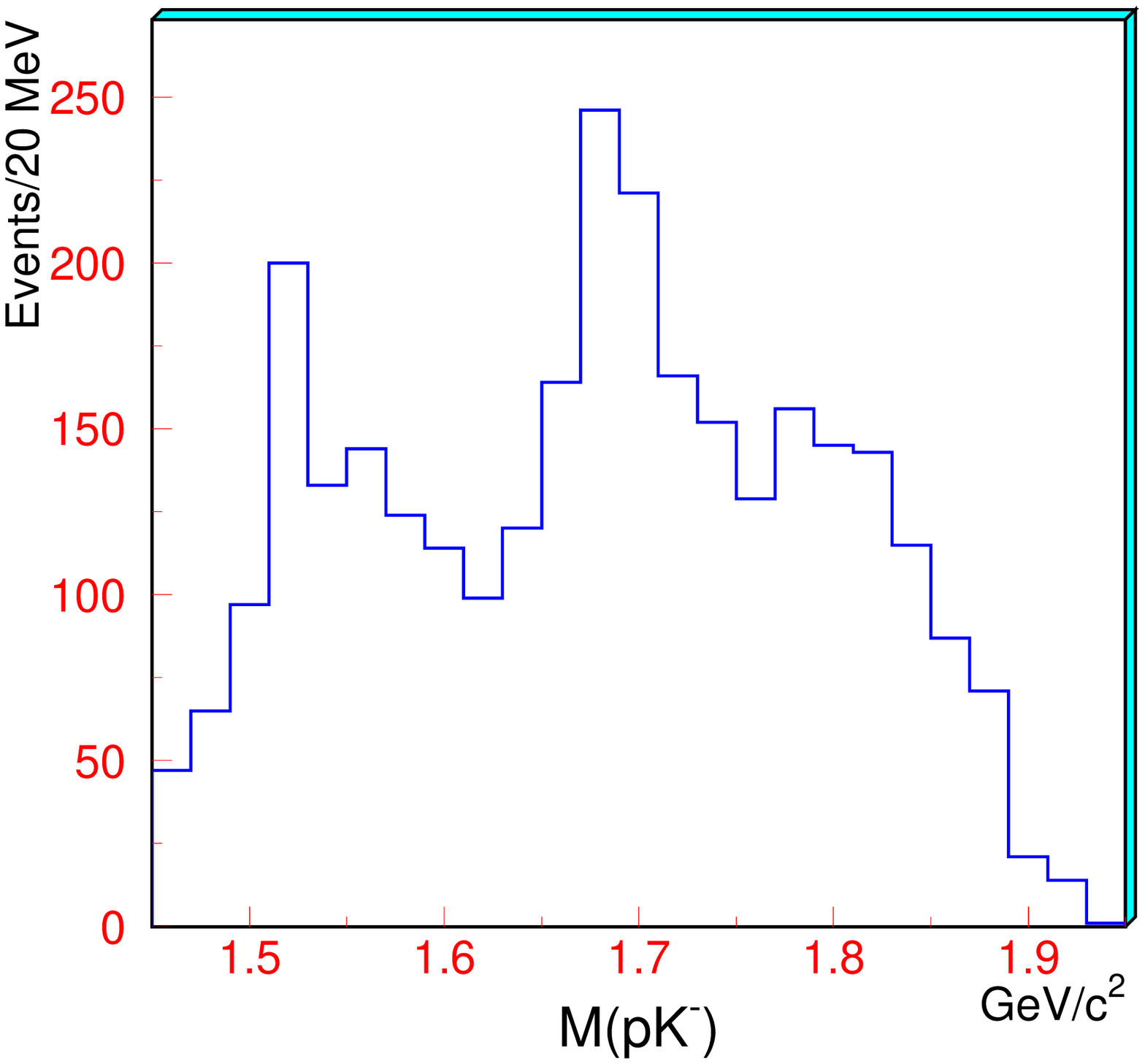,height=8.cm,angle=0,silent=}
\hspace{0.5cm} \psfig{file=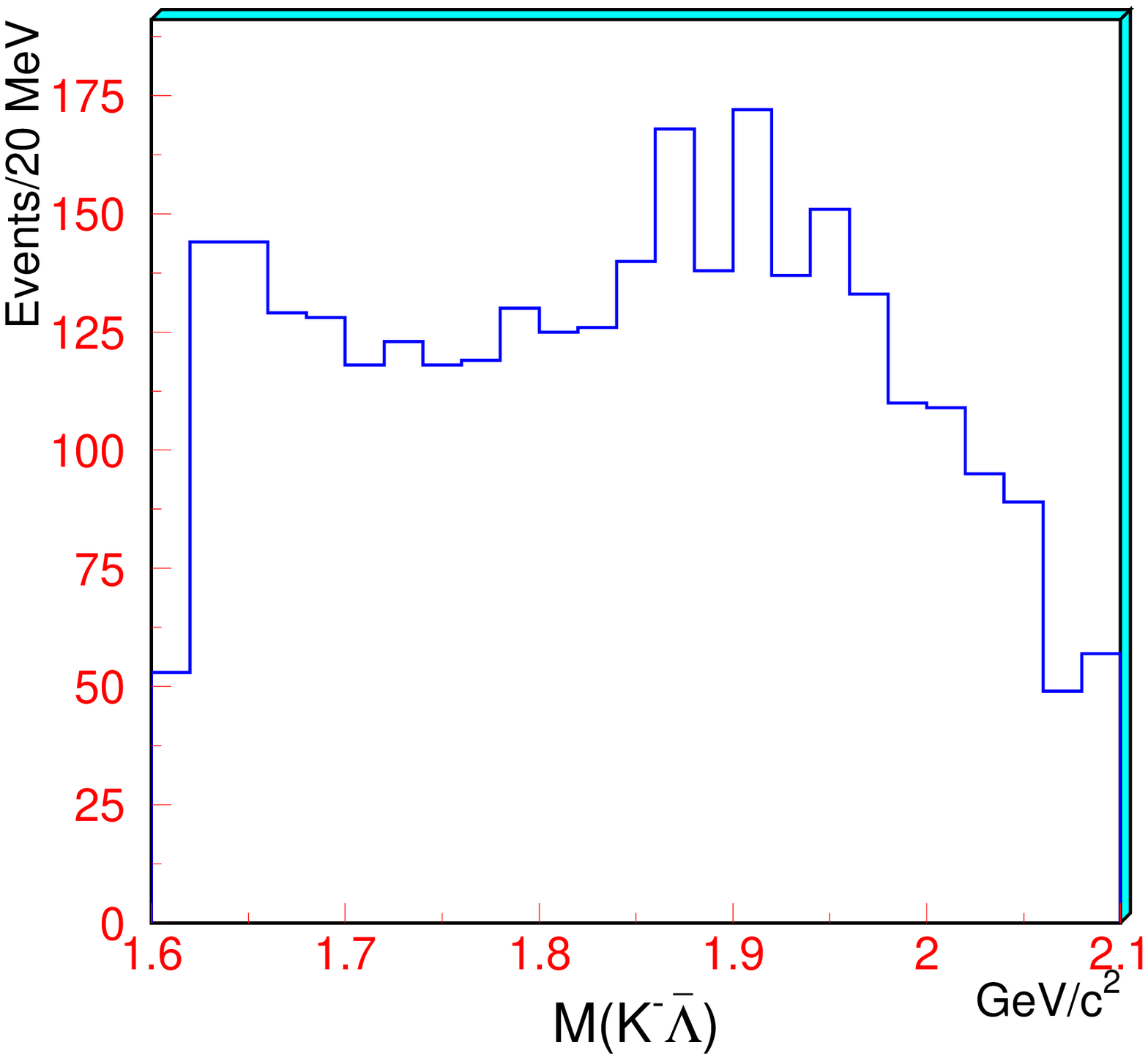,height=8.cm,angle=0,silent=}
}
\vspace{-2.6cm}
\caption{\label{fig4} left: $pK$ invariant mass spectrum for   
$J/\psi\to pK\Lambda$; right: $K\Lambda$ invariant mass
spectrum for
$J/\psi\to pK\Lambda$. Preliminary BESII data}
\end{figure}

With 50 million new $J/\psi$ events collected by BESII of improved
detecting efficiency, we expect to have one order of magnitude more
reconstructed events for each channel. We show in   
Figs.\ref{fig3} and \ref{fig4} preliminary results for $J/\psi\to p\bar
n\pi^-$ \cite{Jixb} and $J/\psi\to pK^-\bar\Lambda + h.c.$ \cite{Yanghx} 
channels, respectively.
For $J/\psi\to p\bar n\pi^-$ channel, proton and $\pi^-$ are detected.
With some cuts of backgrounds, the missing mass spectrum shows a very
clean peak for the missing antineutron with negligible backgrounds; 
The $N\pi$ invariant mass spectrum of 
28,904 reconstructed events from half BESII
data looks similar to the $p\pi$ invariant mass spectrum
for $J/\psi\to p\bar p\pi^0$ as in Fig.~\ref{fig2}, but with much higher
statistics. For $J/\psi\to pK^-\bar\Lambda$ and $\bar pK^+\Lambda$
channels, there are clear $\Lambda^*$ peaks at 1.52 GeV, 1.69 GeV and 1.8
GeV in $pK$ invariant mass spectrum, and $N^*$ peaks near $K\Lambda$
threshold and 1.9 GeV for $K\Lambda$ invariant mass spectrum.

We are also reconstructing $J/\psi\to\bar pp\omega$, $\bar pp\pi^+\pi^-$
and other channels. The $\bar pp\omega$ channel suffers larger background
and the $\bar pp\pi^+\pi^-$ suffers the complication of $\pi\pi$ S-wave
interaction\cite{PDG,Zou3}. 

\section{Partial wave analyses}\label{sec:pwa}

In order to get more useful information about properties of the
baryon resonances, such as their $J^{PC}$ quantum numbers, mass,
width, production and decay rates, etc.,  partial wave analyses (PWA) are
necessary.

\noindent The basic procedure for our partial wave analyses is the 
standard maximum likelihood method:

\noindent (1) construct amplitudes $A_i$ for each i-th possible partial
waves;

\noindent (2) from linear combination of these partial
wave amplitudes, get the total transition probability for each
event as $ w= |\sum_i c_iA_i|^2 $ with $c_i$ as free parameters to be
determined by fitting data;

\noindent (3) maximize the following likelihood function ${\cal L}$ to get
$c_i$ parameters as well as mass and width parameters for the resonances.
$$ {\cal L}=\prod_{n=1}^N\frac{w_{data}}{\int w_{MC}}  , $$
where $N$ is the number of reconstructed data events and $w_{data}$,
$w_{MC}$ are evaluated for data and Monte Carlo events, respectively.

For the construction of partial wave amplitudes, we assume the effective
Lagrangian approach\cite{Nimai,Olsson} with Rarita-Schwinger
formalism\cite{Rarita,Fronsdal,Chung}. In this approach, there are three
basic elements for constructing amplitudes: particle spin wave functions,
propagators and effective vertex couplings; the amplitude can be written  
out by Feynman rules for tree diagrams.

For example, for $J/\psi\to\bar NN^*(3/2+)\to\bar
N(k_1,s_1)N(k_2,s_2)\eta(k_3)$, the amplitude can be constructed as  
$$A_{3/2+}=\bar u(k_2,s_2)k_{2\mu}P^{\mu\nu}_{3/2}(c_1g_{\nu\lambda}
+c_2k_{1\nu}\gamma_\lambda
+c_3k_{1\nu}k_{1\lambda})\gamma_5v(k_1,s_1)\psi^\lambda$$   
where $u(k_2,s_2)$ and $v(k_1,s_1)$ are 1/2-spinor wave functions for   
$N$ and $\bar N$, respectively; $\psi^\lambda$ the spin-1 wave function, 
{\sl i.e.}, polarization vector, for $J/\psi$. The $c_1$, $c_2$ and
$c_3$ terms correspond to three possible couplings for the
$J/\psi\to\bar NN^*(3/2+)$ vertex. The $c_1$, $c_2$ and
$c_3$ can be taken as constant parameters or with some smooth
vertex form factors in them if necessary.
The spin $3/2$ propagator $P^{\mu\nu}_{3/2}$ for $N^*(3/2+)$ is
$$P^{\mu\nu}_{3/2}= \frac{\gamma\cdot p+ M_{N*}}{M^2_{N*}-p^2
-iM_{N*}\Gamma_{N*}}\left[g^{\mu\nu}-\frac{1}{3}\gamma^\mu\gamma^\nu
-\frac{2p^\mu p^\nu}{3M^2_{N*}}
+\frac{p^\mu\gamma^\nu-p^\nu\gamma^\mu}{3M_{N*}}\right]$$
with $p=k_2+k_3$.

Other partial wave amplitudes can be constructed similarly\cite{Liang}.
Programing these amplitudes and maximum likelihood method to fit the data
is straightforward, but very tedious.  Now we are extending the automatic
Feynman Diagram Calculation (FDC) package\cite{Wang} to work for
our partial wave analyses of baryon resonance channels.
Using the extended FDC package, we have performed a partial wave
analysis of the $\bar pp\eta$ channel\cite{Zou1} of BESI data and are now
working on more channels.   

For the $\bar pp\eta$ channel, there is a definite requirement for a
$J^{P}=\frac{1}{2}^-$ component at
$M = 1530\pm 10$ MeV with $\Gamma =95\pm 25$ MeV near the $\eta N$
threshold. In addition, there is an obvious resonance around 1650 MeV
with $J^P=\frac{1}{2}^-$ preferred, $M = 1647\pm 20$ MeV and
$\Gamma = 145^{+80}_{-45}$ MeV.
These two $N^*$ resonances are believed to be the two
well established states, $S_{11}(1535)$ and $S_{11}(1650)$, respectively.
In the higher $p\eta$($\bar{p}\eta$) mass
region, there is a evidence for a structure around 1800 MeV;
with BESI statistics we cannot determine its quantum numbers.

For the $S_{11}(1535)$ propagator, we tried both constant and
energy-dependent width. The Breit-Wigner mass is not sensitive at all to
these two choices. The Breit-Wigner width deffers by 10 MeV for these
two choices: $\Gamma =90\pm 20$ MeV for the case of a constant width
and $\Gamma =100\pm 20$ MeV for the case of an energy-dependent width
assuming $50\%$ to $\eta N$ and  $50\%$ to $\pi N$.
The pole position $(M, \Gamma/2)$ is $(1530, 45)$ MeV and $(1512, 46)$
MeV for assuming constant width and the energy-dependent width,
respectively.    

Preliminary partial wave analysis\cite{Jixb} of BESII data on $\bar
np\pi^-$ channel gives a similar result on $N^*(1535)$ with much smaller
error bars. The outstanding narrow peak around 1.5 GeV in the $\pi N$
invariant mass spectrum demands $N^*(1535)$ to be narrow (width
definitely less than 120 MeV) rather model independently, while PDG gives
a width of $100\sim 250$ MeV.

\section{Future prospects}\label{sec:fu}

BESII just finished data-taking for the 50 million $J/\psi$ events in last
May. We are now working on partial wave analyses of
$J/\psi\to p\bar n\pi^-$, $p\bar p\omega$ channels to study $N^*$
resonances, and $pK^-\bar\Lambda$, $\bar pK^+\Lambda$ channels to study
$\Lambda^*$ resonances as well as $N^*\to\Lambda K$.
As next step, we are going to investigate $\Lambda\bar\Sigma^-\pi^+$,
$pK^-\bar\Sigma^0$ channels to study $\Sigma^*$ resonances; and
$K^-\Lambda\bar\Xi^+$, $K^+\bar\Lambda\Xi^-$ channels to study $\Xi^*$
resonances. These channels are relative easy to be reconstructed by BES.
For example, for $K^-\Lambda\bar\Xi^+$, we can select events containing
$K^-$ and $\Lambda$ with $\Lambda\to p\pi^-$,  then from missing mass
spectrum of $K^-\Lambda$ we should be able to identify the very narrow
$\bar\Xi^+$ peak. We will investigate more complicated channels when we
get more experienced and more manpower. Meanwhile more theoretical
efforts are needed for better partial wave analyses and extraction of
physics from our experimental results\cite{Chiang}.

A major upgrade of the collider to BEPC2 is planned to be finished in
about 4 years. A further two order of magnitude more
statistics is expected to be achieved. Such statistics will enable us
to  perform partial wave analyses of plenty important channels
from not only $J/\Psi$ but also $\Psi'$ decays which will allow us  
to study heavier baryon resonances, {\sl e.g.},  for mass up to 2.36 GeV 
for $\Xi^*$ resonances.
We expect BEPC2 to play a very important unique role in studying excited
nucleons and hyperons, {i.e.}, $N^*$, $\Lambda^*$, $\Sigma^*$ and $\Xi^*$
resonances, and make important discoveries for understanding microscopic
structure of matter.

\section*{Acknowledgments}
We would like to thank Profs. Bill Briscoe and Helmut Haberzettle for
their kind invitation and financial support for our participation of
this very interesting and successful conference. This work is partly
supported by National Science Foundation of China under contract 
Nos. 19991487 and 19905011.

\end{document}